# A Comparative Review Of Internet Protocol Version 4 (IPv4) and Internet Protocol Version 6 (IPv6)

Olabenjo Babatunde[1], Omar Al-Debagy[2]

[1, 2] *Department of Information Systems Engineering*

*Cyprus International University*

*Nicosia, Cyprus*

*Abstract*— **Many computers and devices are becoming more connected to the internet in recent years; the use of the Internet Protocol (IP) has made the connectivity and identification of these devices possible in large scale. In this paper, we will discuss the evolution of Internet Protocol version 4 (IPv4), its features, issues and limitations and how Internet Protocol version 6 (IPv6) tends to solve some of these issues including the differences and transition between these two protocols.**

## I. INTRODUCTION

The Internet Engineering Task Force (IETF) in 1991 decided to create a new version of the Internet Protocol (IP) called Internet Protocol version 6 (IPv6) to replace the old Internet Protocol version 4 (IPv4). This eventually started in 1994[1]. Due to the fact that more and more devices are being connected to the internet, the eventual exhaustion of IPv4 addresses is one of the major reasons why IPv6 was developed.

Currently, the great number of IPv4 users and the huge size of the internet makes an instant migration to IPv6 impossible [1]. Also, because these two protocols can co-exist together due to the auto configuration feature of IPv6, users can enjoy the features of IPv6 and still communicate with IPv4 devices without any problem. We will discuss one of the techniques used by IPv4 to extend the time it will take to exhaust its addresses and also look at some of the techniques and issues involved in migrating to IPv6.

## II. WHAT IS IPV4?

Internet Protocol version 4 (IPv4) has been existing since the early 1980's [3]. It is the forth version of the Internet Protocol (IP) and has been widely used till now. The Internet Protocol is one of the major protocols in TCP/IP. In the OSI model, the protocol works on the Network layer and the major function of the protocol is to identify hosts based on their logical addresses in order to route data between them over the network.

The logical address of a host in a network is the IP address and the IPv4 addressing scheme is what has been used for a while now in identifying hosts in a network. This system is based on a 32-bit logical address [1].

### A. Addressing System

Before the implementation of IPv4, engineers working on ARPANET discussed what the length of an IP address should be; the discussion was whether they should use a 32-bit address or a 128-bit address length. In 1977, a decision was made that a 32-bit length address should be used for IPv4 by Vint Cerf [3]. This was a total of about 4.3 billion addresses and at that time they never foresaw the need for more than that number and this was the beginning of the internet at the time.

IPv4 consists of five classes, A, B, C, D, E. Classes A, B and C have a different bit length for addressing a network host. Class D addresses are reserved for multicasting, while class E addresses are reserved for future use. IPv4 uses a 32 bit addressing which amounts to 4,294,967,296 unique addresses [1]. An example of an IPv4 address is "158.80.164.3", it involves four octets of 8 bits each all resulting to a 32-bit address [5]. In binary form, it would look like 10011110.01010000.10100100.00000011 for the four octets. The table below shows how the classes of IPv4 addresses are assigned including the number of hosts each class has.

TABLE I

| Networks and Hosts per Class | | |
|---|---|---|
| *Network Class* | *Number of Networks* | *Number of Hosts* |
| A | 126 | 16,777,214 |
| B | 16,382 | 65,534 |
| C | 2,097,150 | 254 |

NETWORKS AND HOSTS PER CLASS

The IP addressing system includes a subnet mask which allows us to distinguish between a network address and hosts for example, if we have an IP address of 192.168.0.2 and a subnet mask of 255.255.255.0 the "192.168.0" identifies the network which is a class C network address and the last octet in decimal "2" identifies the host.

### B. Why are we running out of IPv4 addresses?

In recent years, the use and production of more handheld devices such as mobile phones and tablet including the use of more computers all connecting to the internet have increased the demand for more IPv4 addresses. On the 3rd of February 2011, the Internet Corporation for Assigned Names and Numbers (ICANN) released the last block of the IPv4 addresses [3]. This evidently shows that we are running out of IPv4 addresses, which means it would be difficult to allocate

IP addresses to new or expanding companies in the future. Just because we are running out of IPv4 addresses, several methods have been employed to increase the time it would take before we completely run out of IPv4 addresses.

*C. IPv4 shortage workaround*

Although we are currently running out of IPv4 addresses, some technologies have been employed to work around this issue. The most common are Network Address Translation (NAT), Classless Inter-Domain Routing (CIDR), and dynamic IPv4 address assignment (DHCP) or Dynamic Host Configuration Protocol [4]. Network Address Translation (NAT) has been the most popular of these technologies and it has helped shift further the time it would take before IPv4 addresses are exhausted [3].

*D. Network Address Translation (NAT)*

To reduce the number of addresses needed for an organization in order to save IP addresses, Service Providers (SP) try to reuse address blocks by using multiple layers of Network Address Translation (NAT) [3]. Network Address Translation allows a single device, such as a router, to act as an agent between the Internet "public network" and a local "private" network. This allows a single IP address to represent an entire group of computers or devices.

NAT translates public IP addresses into private IP networks; private networks are networks that cannot access the internet [4]. They consists of the following IP addresses for each class of network; For Class A, network 10 is used, for the Class B, networks 172.16 to 172.31 is used and for Class C, networks 192.168.0 to 192.168.255 is used in the private network [4].

With the use of NAT, Internet Service Providers (ISPs) do not need to assign individual IP addresses to each customers device, they only need to assign a single IP address to the customer and then this address can be used via NAT to provide more private IP addresses to other devices the customer may have thus enabling more IP addresses to be available to more customers.

*E. Shortcomings of IPv4*

Internet Protocol version 4 (IPv4) has a few shortcomings stated below;

*1) Address Space:* Due to the number of increasing devices connected to the internet, the IPv4 addresses can only accommodate about 4 billion hosts. This is a significant limitation as more and more devices are getting online.

*2) Security:* IPv4 does not provide any security like authenticating packets [5] when they are transmitted or encryption of the data.

*3) Network Congestion:* Due to the broadcast feature in IPv4 network devices can become congested and overloaded since packets are sent to all addresses in the network.

*4) Packet Loss:* IPv4 contains a Time to Live (TTL) [5] field in the header of an IP and this set the expiry time for the datagram. If the data was unable to get to the destination on time, it will expire and will be requested again from the receiving computer. This delay and multiple requests can cause packets to be lost and this is not good for real-time data like VOIP or video streaming.

*5) Data Priority:* Due to the fact that IPv4 cannot recognize the kind of data being transmitted its difficult for the protocol to prioritize transmission high priority data like video streaming and others.

### III. WHAT IS IPV6?

Internet Protocol (IPv6 or IPng) is the next generation of IP and it is the successor of IP version 4 which is widely used nowadays. The development of IPv6 started in 1991 and was completed in 1997 by the Internet Engineering Task Force (IETF), and was officially used in 2004 when ICANN added IPv6 addresses to its DNS server [2].

Data transfers between hosts in packets across networks, these packets require addressing schemes. Using IPv4 and IPv6 these packets can identify their sources and also find their destinations. Every device on the Internet needs an IP address to communicate with other devices, and the growth of the Internet led to a need for a new alternative for IPv4, because IPv4 cannot provide the needed number of IP address around the world [6].

The address space in IPv6 is much larger than the address space of IPv4, and it went from 32 bits to 128 bits; in other words, it went from 4 billion addresses to 340 trillion trillion trillion of unique address [2]. IPv6 is designed to provide unique addresses for everyone on earth. This expansion in address space will not just provide more unique address but it will also make routing easier and cleaner because of its hierarchical addressing and simpler IP header [2].

The IPv6 addressing structure is designed to provide compatibility with existing IPv4 networks and allows the existence of both networks. IPv6 does not only solve the problem of shortage that IPv4 is causing, but it is also enhances and improves some of the features that IPv4 has [4].

IPv6 uses 128 bits addressing format that is represented by 16-bit hexadecimal number fields separated by colons ":". Using this format makes IPv6 less messy and error-free. Here is an example of an IPv6 address [2]:

2031:0000:130F:0000:0000:09C0:876A:130B

Additionally, this address can be shortened using some rules like compressing the block of zeros to a single zero like this [2]:

2031:0:130F:0:0:9C0:876A:130B or 0000=0

Also, successive fields of zero can be represented by double colons "::", but it is only allowed once to use a double colon, so the above example will be shortened to this:[2].

2031:0:130F::9C0:876A:130B

*A. Solving problems in IPv4*

The main reason behind adapting IPv6 is the exhaustion of available IPv4 address space. IPv6 has many new features and some critical improvements over IPv4. Due to the rapid growth of Internet users in the past ten years and the ongoing growth, IPv4 is being exhausted and its address space will not be able to satisfy the huge number of Internet users and does not provide the geographical needs for the expansion of the Internet. Also, the emergence and the growth of mobile devices like smart phones and tablets made the problem of IPv4 more urging than before.

IPv6 will provide larger address space for more reach and scalability and this will almost provide unlimited number of IP addresses and more efficient techniques for routing, so these features will provide global addresses for each network device and will enable end-to-end reachability and better network performance.

In addition to solving the problems of IPv4, IPv6 will provide networks and IT professionals with a list of benefits and features which are listed below [4]:

- Larger address space.
- Hierarchical network architecture.
- Simpler header format which will make packet handling more efficient.
- There will be no need for network address translation (NAT) and application's layered gateway (ALG).
- Built-in security with IPSec implementation.
- Auto-configuration and plug-and-play support.
- Expanding the number of multicast addresses.
- Improved support for Mobile IP and Mobile Computing Devices.

*B. IPv6 Migration Issues and Techniques*

The migration from IPv4 to IPv6 has already started, but before launching IPv6 the network infrastructure should be upgraded in order to support services and software. But there are some issues should be taken care of before migrating from IPv4 to IPv6, and these issues are:

*1) Infrastructure Issues:* Many of the protocols and technologies should be redesigned in order to support IPv6 including DHCP, OSPF, and RIP, BGP, ARP, TCP/IP and others [6].

*2) Tunnelling Issues:* Without any change in applications, IPv6 can be used in an existing network by using IPv6 over IPv4 tunnelling for connecting the IPv4 nodes to the backbone network. But tunnelling has very less throughput and it needs network managers to configure the tunnel end points information, which is a time consuming process [6].

*3) Financial Issues:* Migrating from IPv4 to IPv6 requires purchase of new network devices that support IPv6 like switches, routers, and others devices, which will make enterprises and companies invest more money on the migration process [6].

*4) Security Issues:* IPv6 have not been used widely and has not been tested properly. So, the security level of IPv6 is still vague [6].

There are many techniques and mechanisms that will make the transition from IPv4 to IPv6 an interoperable operation. Some of these techniques are:

*1) Dual Stack:* IPv6 is an upgrade for IPv4, so IPv6 inherits some features from IPv4; therefore it is relatively easy to create a network stack that supports both IPv4 and IPv6. Such an implementation is called a dual stack. Most of the new network devices and software which provide IPv6 have implemented dual stack mechanisms [1].

*2) Tunnelling:* If we want an internet that is fully depended on IPv6, the implemented IPv6 networks and hosts must be able to transfer IPv6 packets through the existing IPv4 infrastructure. So this can be done using a technique called tunnelling, which consists of encapsulating IPv6 packets within IPv4, which means using IPv4 as a link layer for IPv6 [4].

*3) Addressing simplicity:* Address simplicity allows a router or host being updated to IPv6 to continue to use IPv4 address more like automatic tunnelling which are simple mechanisms that provides IPv6 connectivity between separated dual-stack hosts, routers, or both [1].

*4) Proxying and Translation:* This technique is used when an IPv6 device tries to access an IPv4 service like a web server, so there should be some kind of translation between these two end-points in order to connect to each other. Therefore, the most reliable way of translation is the use of dual-stack application-layer proxy, or in other words a web proxy [4].

## IV. COMPARING IPV4 WITH IPV6

Although NAT in IPv4 helped reduce the number of public IP addresses needed in an organization, NAT still has some security and performance issues. NAT being good for client server communication such as email and web has issues when it comes to peer-peer communication [4]. IPv6 provides an end to end network connection which is a peer-peer system used in applications like VOIP. It also has an auto configuration system that allows clients to communicate independently without any need for a manual setup and also makes use of IPSec compulsorily in all its communication. This make IPv6 more secure than IPv4 [5].

Also, because IPv4 has fewer addresses than IPv6, this will require the use of proxies and other forms of network mapping, thereby increasing the risk in packet sniffing through proxies but IPv6 contains more address space thereby reducing the use of proxies and ultimately increasing the level of security on the network.

The table below shows some of the key differences between IPv4 and IPv6.

TABLE III
DIFFERENCES BETWEEN IPv4 AND IPv6

| Differences between IPv4 and IPv6 | |
|---|---|
| *IPv4* | *IPv6* |
| IPv4 uses a 32 bits address space | IPv6 uses a 128 bits address space |
| Must support DHCP or be configured manually | Does not require DHCP or manual configuration, it supports stateless auto configuration[1] |
| IPSec is not compulsory | IPSec is compulsory |
| Broadcasts sends traffic to all hosts on a subnet | There are no broadcasts instead multicasts is used thereby reducing broadcast floods found in IPv4 |
| The IP header has a variable length of 20-60 bytes depending on the IP header options | IP header has a fixed length of 40 bytes and there are no IP header options available |

## V. CONCLUSION

The future of the internet is IPv6 but migrating from IPv4 to IPv6 is a gradual process and may take several years before we completely migrate. Several techniques being used to maintain interconnectivity between the two protocols will allow these protocols to coexist without issues as more companies migrate to the new protocol.